\documentstyle[multicol,prb,aps,epsfig]{revtex}

\begin{document}
\title{Fe/V and Fe/Co (001) superlattices: growth, anisotropy, magnetisation and 
magnetoresistance}

\author{P. Nordblad, A. Broddefalk, R. Mathieu}
\address {Department of Materials Science, Uppsala University, Box
534, SE-751 21, Uppsala, Sweden}
\author{P. Blomqvist, O. Eriksson, R. W\"appling }
\address {Department of Physics, Uppsala University, Box
530, SE-751 21, Uppsala, Sweden}

\date{\today}
\maketitle

\begin{abstract}
Some physical properties of bcc Fe/V and Fe/Co (001) superlattices are reviewed. 
The dependence of the magnetic anisotropy on the in-plane strain introduced by 
the lattice mismatch between Fe and V is measured and compared to a theoretical 
derivation. The dependence of the magnetic anisotropy (and 
saturation magnetisation) 
on the layer thickness ratio Fe/Co is measured and a value for the anisotropy of 
bcc Co is derived from extrapolation. 
The interlayer exchange coupling of Fe/V superlattices is studied as a function of the 
layer thickness V (constant Fe thickness) and layer thickness of Fe (constant V 
thickness). A region of antiferromagnetic coupling and GMR is found for V 
thicknesses 12-14 monolayers. However, surprisingly, a 'cutoff' of the 
antiferromagnetic coupling and GMR is found when the iron layer thickness 
exceeds about 10 monolayers.    

\end{abstract}

\pacs{PACS numbers: 75.70.-i,75.30.Gw, 75.70.Pa
\\Keywords: Magnetic superlattices, anisotropy, magnetisation, GMR, Fe, Co
\\Corresponding author: per.nordblad@angstrom.uu.se}

\begin{multicols}{2}

\section{INTRODUCTION}
The magnetic properties of the transition elements are critically dependent on fine 
details of the electronic structure of the d-electrons. Thus, any changes of the 
symmetry or interatomic distances as well as restrictions in dimensionality of the 
system could introduce striking changes of the magnetic properties. Artificial, but 
controlled, changes of symmetry, inter atomic distances and dimensionality are 
readily introduced in Fe and Co when they are grown on single crystalline substrates 
in the form of thin films or 
superlattices. Such layered structures allow systematic studies of the dependence of 
major magnetic properties on e.g. controlled magnitudes of the induced strains.
In this article we review results from some recent studies of the magnetic 
properties of (001) bcc iron and cobalt. \cite{MAE,GMR,FeCo}

\section{Growth}
Bcc Fe/V (001) and Fe/Co (001) superlattices were grown on 
MgO (001) substrates by UHV dc sputtering \cite{PIs,PBl}. The lattice constants of 
the involved materials are: Fe (2.87 A), V (3.03 A), MgO (4.21 A) and 
bcc Co (2.82 A) (bcc Co does not exist naturally, but the lattice parameter is estimated 
from extrapolation from values for Fe(Co) alloys). The lattice mismatch between 
Fe and V is small enough to allow epitaxial growth and still large enough to 
introduce relevant strain in the individual Fe layers. The in-plane strain introduced 
in the samples due to the lattice mismatch causes a symmetry breaking and the 
true crystallographic structure of the samples becomes body centred tetragonal (bct). 
The superlattices grow with an [100] direction of Fe (V or Co) 
along an [110] direction of MgO, 
yielding a lattice mismatch of -4\% for iron and 2\% for V. The samples were grown under
optimal conditions \cite{PIs,PBl} and the crystallographic properties of the samples
were measured by different X-ray techniques certifying the crystallographic quality as 
well as giving measures of the lattice parameters of the superlattices and the 
individual layer thicknesses.

\section{Magnetic and transport properties}
The magnetic and transport properties of the samples were investigated using a 
QD MPMS5 SQUID-magnetometer and an Oxford Maglab 2000 system. 

\subsection{Fe/V - magnetic anisotropy energy}
The in-plane magnetic anisotropy energy (MAE) of different Fe/V samples 
was derived from measurements 
of magnetisation curves along the in-plane [100] and [110] directions of the films. 
\cite{MAE} 
No uniaxial contribution to the MAE was found in any of the samples, 
which also certifies a good bct crystallographic structure of the superlattices. 
The quite significant lattice mismatch between
Fe and V allows studies of the anisotropy as a function of in-plane lattice
strain of Fe. Samples of constant iron layer thickness (15 monolayers (ML)), but varying 
V thickness (1-12 ML)
are studied, the in-plane lattice parameter of the samples increases with increasing 
thickness of 
the V layers. Experimental average in and out of plane lattice parameters were derived
from X-ray diffraction and were 
complemented by theoretical calculations of the common
in-plane and individual out of plane lattice parameters using tabulated 
elastic constants for Fe and V. The measured and calculated average lattice 
parameters agreed fairly well and the theoretically derived 
individual in and out of plane lattice parameters could then safely be used for 
theoretical calculations of the magnetic properties. 
 
\begin{figure}
\centerline{\epsfig
{figure=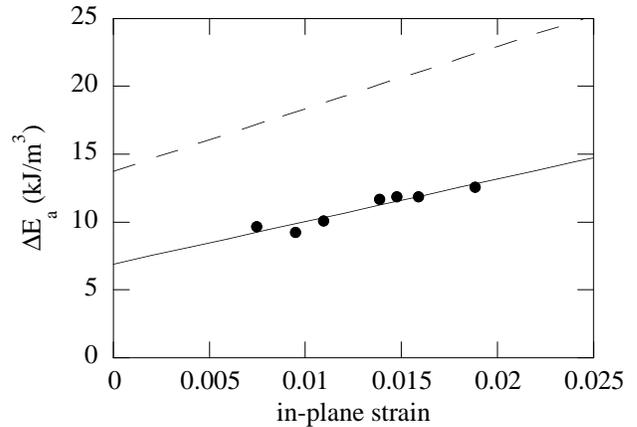}}
\caption{
\label{fig:MAE}
{The magnetocrystalline anisotropy energy difference between [110]
and [100] of Fe/V (001) at T=10 K. The marks and the solid line denotes the 
experimental and the dashed line the 
calculated results, from ref. [1].}
}
\end{figure}

The saturation magnetisation of the superlattices was found to be reduced in comparison 
with bulk iron, amounting to 1.9 $\mu$$_B$ compared to 2.2 $\mu$$_B$ for bulk iron. The 
reduced magnitude can partly be explained by an induced weak antiferromagnetically 
aligned magnetic moment in the vanadium interface layers.  Fig. \ref{fig:MAE} shows 
the measured magnetic anisotropy as a function of lattice strain at 10K. 
The corresponding theoretically calculated 
values for strained bulk iron are also included in the 
figure. These values have been
derived using the magnetoelastic constants of iron and the known anisotropy for 
unstrained bulk iron. The calculated strain dependence of the MAE is remarkably 
similar to that experimentally observed, 
but there is a significant downward shift of the measured 
MAE compared to that of bulk iron, which can be assigned to a surface contribution 
favouring [110] as the easy direction at the Fe/V interfaces. 

When discussing anisotropy of Fe/V superlattices it should be noted that 
superlattices grown to form (110)
planes do show a totally dominating uniaxial in-plane anisotropy of a magnitude
that is comparable to the anisotropy of hexagonal cobalt, although the in-plane symmetry
breaking is due to quite modest strains in the films.\cite{110} One interesting 
future study of the current (001) 
Fe/V system would thus be to investigate the anisotropy of these superlattices by e.g. 
ferromagnetic resonance (FMR) to get the full picture of both the in-plane and the 
out of plane anisotropy and the influence of the bcc to bct symmetry breaking.  

\subsection{Fe/V - GMR}
\begin{figure}
\centerline{\epsfig
{figure=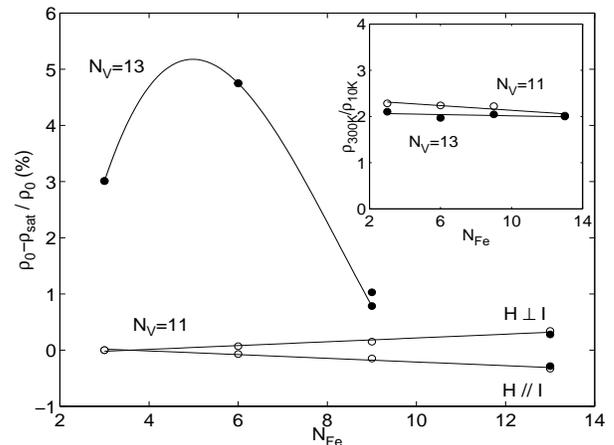,height=6cm,width=8cm}}
\caption{
\label{fig:GMR-MR}
{The magnetoresistance vs Fe thickness of Fe/V (001)for N$_V$ = 11 and 13 ML
at T=10 K, from ref. [2]. }
}
\end{figure}

The Fe/V (001) system has also been studied 
as to the magnetoresistance and interlayer 
coupling strength as a function of the V or the Fe layer thickness. \cite{GMR,Gb1,Gb2} 
In the investigated range of V thicknesses (2-16 ML), the interlayer coupling shows
one antiferromagnetic maximum centred at V $\approx$ 13 ML. Fig. \ref{fig:GMR-MR} 
shows the dependence of the 
magnetoresistance for two sample series with vanadium thicknesses of 11 and 13 ML and 
varying Fe thickness.\cite{GMR} The 11 ML V samples show only an anisotropic 
magnetoresistance (AMR) of increasing magnitude with increasing iron layer thickness. 
However, the 13 ML V samples show GMR for thin to intermediately thick Fe layers, 
but for thicker Fe layers the GMR (and the antiferromagnetic (AF) inter-layer 
coupling) disappears. The GMR effect is superposed by a weak AMR of 
similar magnitude as observed in the non-AF coupled superlattices. The most surprising 
feature is the appearance of a cut-off of the GMR occurring for iron 
thicknesses larger than about 10 ML, leaving only an AMR effect in the samples. 
This apparent cut-off of the AF-coupling finds no support in current
theory \cite{dGMR} and the physical  origin of it is thus not yet understood.  
However, the fact that the current samples consist of 30 bilayer repetitions
causes some magnitude mismatch between the magnetisation of 
the AF coupled Fe layers. This yields a net magnetic moment that could be large enough 
to interfere with the AF coupling and destroy the regular AF magnetic structure 
throughout the sample that is necessary to give the GMR effect and a measurable 
AF-coupling strength. 
 
\subsection{Fe/Co - magnetisation and magnetic anisotropy}
\begin{figure}
\centerline{\epsfig
{figure=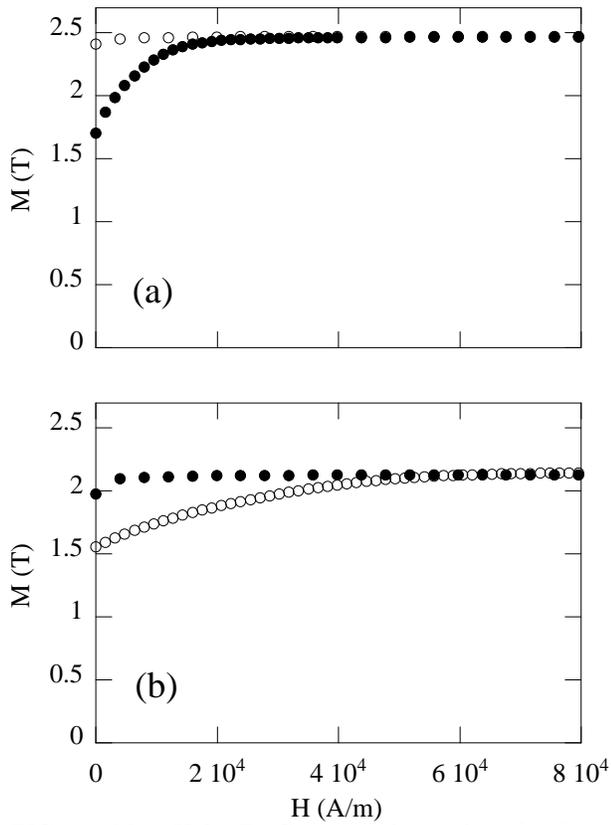}}
\caption{
\label{fig:MvsH}
{M vs H for Fe/Co 6/2 and 2/6 along [100] (open circles) and [110] 
(filled circles) at T=10 K, from ref. [3].}
}
\end{figure}

\begin{figure}
\centerline{\epsfig
{figure=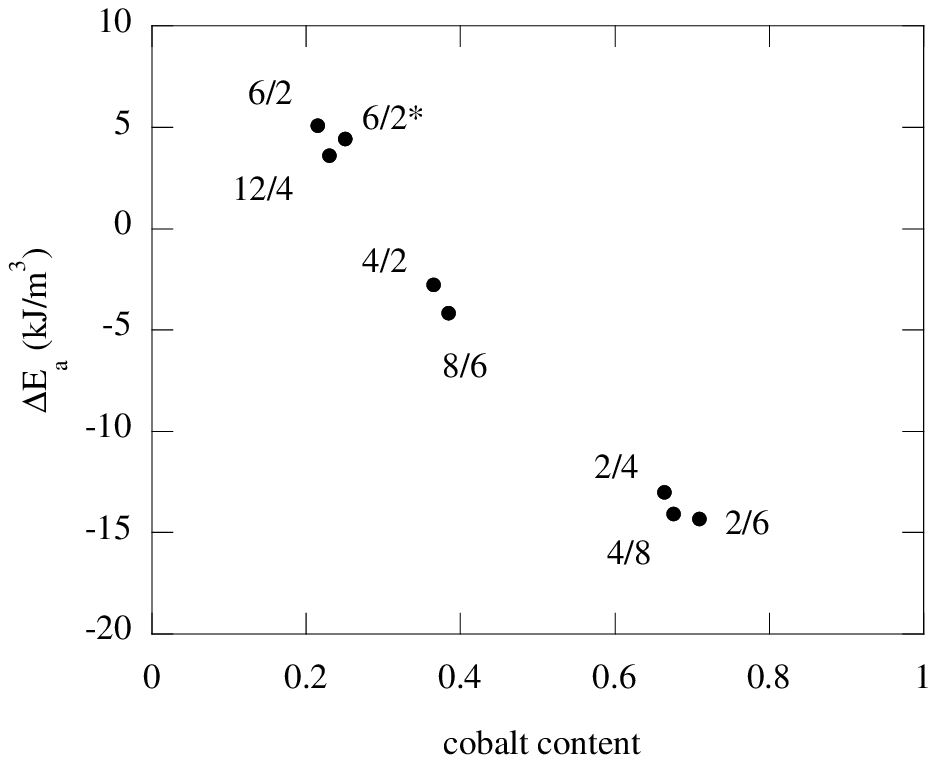}}
\caption{
\label{fig:MAECo}
{MAE vs Co content for the different Fe/Co (001) samples at T=10 K, from ref.  
[3].}
}
\end{figure}

Cobalt does not naturally crystallise in the bcc structure, it is however possible to grow
bcc Co as a superlattice intervened with Fe. These bcc Fe/Co superlattices show 
intriguing magnetic properties.\cite{FeCo} The magnetisation curves for an iron rich and
a Co rich superlattice with bilayer thicknesses 6/2 and 2/6 are shown in 
Fig. \ref{fig:MvsH}. 
There is a change 
of sign of the anisotropy energy as the system becomes richer on cobalt. 
In Fig. \ref{fig:MAECo} the magnetic anisotropy energy (MAE) between the [100] and [110] 
directions is plotted vs. Co content in Fe/Co samples of bilayer thicknesses (x/y) 
as indicated in the figure. There is a closely linear dependence of the 
anisotropy energy on concentration. An extrapolation to pure Co gives a value
of the MAE that has opposite sign and about twice the magnitude
of the MAE of bulk iron. However, one should keep in mind that there can be
a significant contribution from the interface regions (cf the results above on 
the Fe/V system) that may move the MAE curve
downward as compared to what would be observed for pure bulk iron and pure but 
artificial bulk bcc cobalt.

\begin{figure}
\centerline{\epsfig
{figure=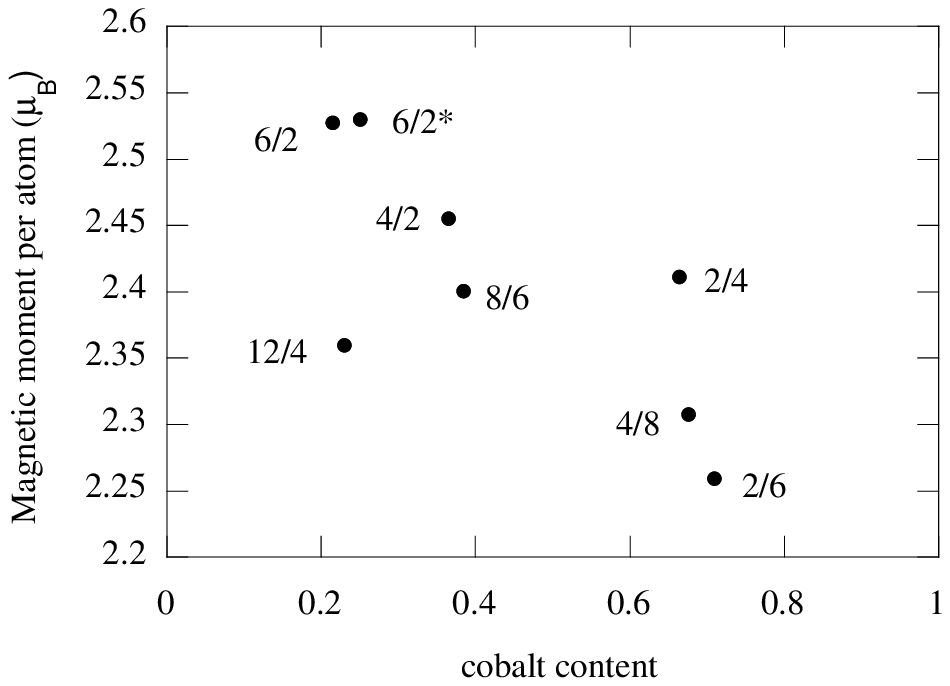}}
\caption{
\label{fig:MsCo}
{M$_s$ vs Co content of the different Fe/Co (001) samples at T=10 K, from ref.  
[3].}
}
\end{figure}

The saturation magnetisation of the different samples are shown in 
Fig. \ref{fig:MsCo}. The
values are given in $\mu$$_B$/atom. There are several features to note in this 
figure. The most striking is that the magnitude of the magnetisation is 
significantly larger than that of corresponding Fe/Co alloys, i.e. there is
an interface enhancement of the magnetic moments. The current data gives a
possibility to phenomenologically assign a specific value of the magnetic moments 
of Fe and Co in the atomic  
layers at the interface and deeper into the magnetic layers. Doing this, one 
finds a significant
enhancement of the iron moments at the interfaces, and also up to some 4 layers
away from the interface. To account for the observed behaviour one has also to
introduce an oscillating nature of the magnetisation in the layers having the 
largest magnitude at the 
interface, a value close to the bulk value one atomic layer away and then again 
significantly enhanced values
in the following two atomic layers to finally settle at the bulk iron value 
deep into a thick Fe-layer. It was also found necessary to introduce a significantly 
enhanced Co moment
at the interface. The enhanced value of the magnetisation in a superlattice 
as compared to
the corresponding bulk alloy does ask for further experimental and theoretical 
studies, since current
theory does not predict such a big change and basically no enhancement of the
Co moment at the interfaces. \cite{Nik}

\section{conclusions}
The magnetism of the transition metals (e.g. Fe and Co) depends critically on the 
symmetry and volume of the lattice. These parameters are readily varied in 
theoretical calculations and the magnetism of the elements can be mapped out for
'any' artificial crystallographic structure. One possibility to check these
predictions and to tailor the magnetic properties of a magnetic material is given by 
growing magnetic superlattices containing two different transition 
metals. We have here briefly reviewed some recent experimental results on two
different superlattice systems Fe/V (001) \cite{MAE,GMR} and Fe/Co (001) 
\cite{FeCo} that are and have been extensively studied at the 
Angstrom Laboratory in Uppsala.\cite{PIs,PBl,brodden}

\section{ACKNOWLEDGEMENTS}

Financial support from the Swedish Research Council (VR) and the Swedish Foundation 
for Strategic Research (SSF) is acknowledged.

\end{multicols}
\end{document}